\documentclass[preprint]{raa}  
\usepackage{graphicx,times}
\usepackage{natbib}
\usepackage{amssymb,amsmath}
\bibpunct{(}{)}{;}{a}{}{,}
\def\hi{H{\textsc{i}}}
\def\kms{km~s$^{-1}$}

\usepackage{color}
\usepackage{multirow}
\usepackage{rotating}
\usepackage{amsmath} 
\usepackage{amsmath}
\usepackage{natbib}
\usepackage{graphicx}
\usepackage{mathrsfs}
\usepackage{color}
\usepackage{booktabs}
\usepackage{lipsum}
\usepackage{epstopdf}
\usepackage{rotating}
\usepackage{longtable}

\usepackage[pagebackref=true]{hyperref}

\begin{document}

   \title{ First Search for Pulsed CH Maser Emission Stimulated by a Pulsar
   }

 \volnopage{ {\bf 20XX} Vol.\ {\bf X} No. {\bf XX}, 000--000}
   \setcounter{page}{1}

   \author{Mengting Liu 
   \inst{1}, Di Li\inst{2,3},  J.\ R.\ Dawson\inst{4,5}, Joel M.\ Weisberg\inst{6}, George Hobbs\inst{4}, Ningyu Tang \inst{7},   Gan Luo \inst{8}, Duo Xu\inst{9}, Donghui Quan \inst{1}
   }

   \institute{ Research Center for Astronomical Computing, Zhejiang Laboratory, Hangzhou 311100, China; {\it liumengting@nao.cas.cn; donghui.quan@zhejianglab.org}\\
        \and
             Department of Astronomy, Tsinghua University, Beijing 100084, China; {\it dili@tsinghua.edu.cn}\\
	\and
	 National Astronomical Observatories, Chinese Academy of Sciences, 20A Datun Road, Beijing 100101, China\\
\and 
CSIRO Astronomy $\&$ Space Science, Australia Telescope National Facility, P.O. Box 76, Epping, NSW 1710, Australia\\
\and
School of Mathematical and Physical Sciences and Astrophysics and Space Technologies Research Centre, Macquarie University, NSW 2109, Australia\\
\and
Department of Physics and Astronomy, Carleton College, Northfield, MN 55057\\
\and
Department of Physics, Anhui Normal University, Wuhu, Anhui 241002, People's Republic of China\\
\and
Institut de Radioastronomie Millimetrique, 300 rue de la Piscine, 38400, Saint-Martin d\'Heres, France\\
\and
Canadian Institute for Theoretical Astrophysics, University of Toronto, 60 St. George Street, Toronto, ON M5S 3H8, Canada\\
\vs \no
   {\small Received 20XX Month Day; accepted 20XX Month Day}
}

\abstract{We present the first search for pulsed CH maser emission potentially stimulated by PSR~J1644$-$4559, conducted using the ultra-wide-bandwidth low-frequency receiver on Murriyang, CSIRO's Parkes Radio Telescope. 
Observations targeted three CH $\Lambda$-doublet transitions at 3264, 3335, and 3349 MHz, with a variability timescale of 78 ms. 
We detected ten CH emission features at 3335 and 3349 MHz, and seven features at 3264 MHz, during both pulsar-ON and pulsar-OFF phases. 
The observed velocities align with the OH emission and absorption reported by a previous study, suggesting a close spatial association between CH and OH molecules. 
The derived column densities for CH clouds within the Parkes beam range from $0.05$ to $9.8 \times 10^{13}$ cm$^{-2}$, indicating that these clouds are likely in diffuse and translucent states. 
Upper limits for CH column densities within the pulsar beam ranged from $0.3$ to $9.8 \times 10^{13}$ cm$^{-2}$. Comparison of these column densities suggests that CH clouds may exhibit clumpiness and substructure.
No significant stimulated emission feature was detected in the optical depth spectra. 
Additionally, as part of our search for pulsed stimulated emission, we investigated the potential CH absorption of the pulsar signal and found none, in agreement with astrophysical expectations. 
The upper limits for the potential maser amplification factors towards PSR~J1644$-$4559 at 3264, 3335, and 3349 MHz are 1.014, 1.009, and 1.009, respectively. 
This study demonstrates the feasibility of detecting pulsed CH maser emission in the interstellar medium stimulated by pulsar photons. 
}

   \authorrunning{Liu et al.}            
   \titlerunning{Pulsed CH Maser Search towards PSR~J1644$-$4559}  
   \maketitle

%
\section{Introduction}           
\label{sect:intro}

The only reported instance of natural, fast-switching stimulated emission was observed toward PSR~J1644$-$4559 (also known as B1641$-$45) in the hydroxyl radical (OH) 1720 MHz transition \citep{Weisberg_etal_2005}. 
Utilizing gated-mode observations at Murriyang, CSIRO's Parkes Radio Telescope, the pulsar data were binned every 14 ms to produce pulsar-ON and pulsar-OFF spectra. 
For the 455 ms period of J1644$-$4559, this time resolution corresponds to 1/32 of the full phase. Since the flux change appeared instantaneous at this time interval, \citet{Weisberg_etal_2005} claimed such pulsar-stimulated emission to be the shortest fluctuation observed in any interstellar masers.

PSR~J1644$-$4559 is the only source, out of a sample of 18 pulsars chosen by \citet{Weisberg_etal_2005} for OH absorption studies, to exhibit pulsar-stimulated emission; demonstrating the uncommon nature of this phenomenon. 
The Parkes OH spectra of J1644$-$4559 are complex, exhibiting a mix of absorption, emission (stimulated and not), and conjugate relations between OH transitions. 
The richness of ISM structures and conditions within the telescope beam present both challenges and incentives for further investigation of this intriguing phenomenon. 
The pulsar-OFF spectrum samples material within the entire telescope beam, while the pulsar-ON spectrum probes a pencil-beam 
volume through the ISM; thereby providing a natural differential tracer of millisecond time scale variability of spectral lines.

\begin{figure*}
\centering
\includegraphics[width=0.45\linewidth]{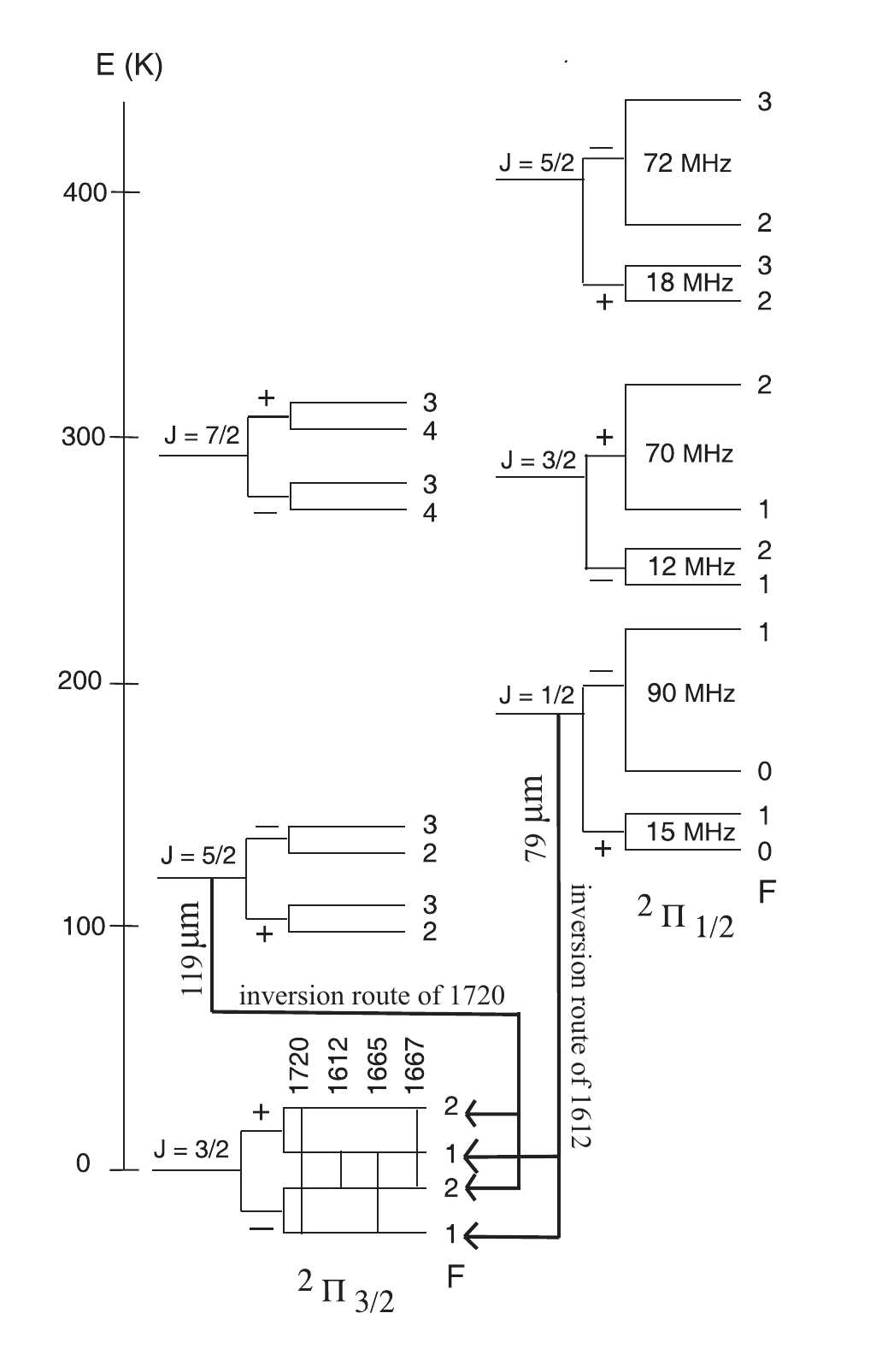}
\includegraphics[width=0.45\linewidth]{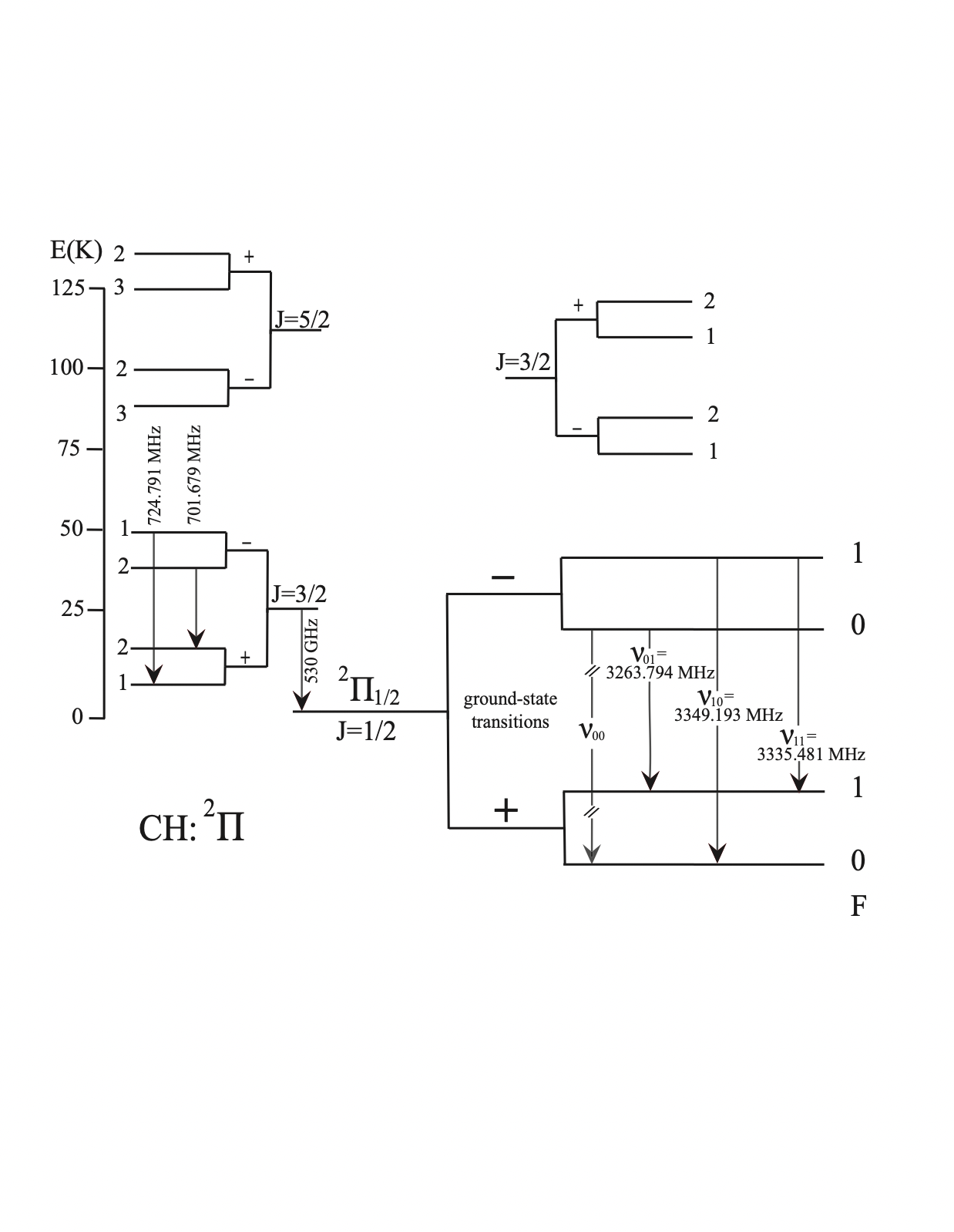}
\caption{OH (left) and CH (right) energy levels diagrams adapted from \citet{Xu_2016_OH} and \citet{Xu_2016_CH} (not shown to scale). The observed OH transitions from \citet{Weisberg_etal_2005} correspond to the $^{2}\Pi_{3/2}, J=3/2$ ground state, while the CH transitions in this work correspond to the $^{2}\Pi_{1/2}, J=1/2$ ground state. 
}
\label{fig:OH_CH_eng}
\end{figure*}

The $\Lambda$-doubling transitions of CH around 3.3 GHz are similar in many ways to those of OH in the L-band, as shown in Figure~\ref{fig:OH_CH_eng}. 
The $\Lambda$-doubling energy levels lie in 3.3 GHz bands with modest A-coefficients, while the rotational transitions lie in the submillimeter to far-infrared bands. 
Such configurations can lead to an excess population in the upper levels of the $\Lambda$-doubling transitions, resulting in phenomena like masers \citep{1992ASSL..170.....E}. 
In fact, the ground-state transitions of CH tend to be anomalously excited \citep{Turner_1988_CH,Ningyu_2021_CH,Jacob_2021_CH,Jacob_CH_2023}. 
While \hi\ has been studied in absorption towards pulsars for decades (e.g. \citealp{Weisberg_etal_1979,Clifton_1988_HI_PSR,Frail_1994_PSR_HI,Simon_2003_HI,Snez_HI_2010,Liu_HI_2021}), OH has only been clearly detected toward three pulsars out of about 30 targets searched (B$1849+00$: \citealp{Snez_OH_2003}; B$1641-45$: \citealp{Weisberg_etal_2005}; B$1718-35$: \citealp{Minter_2008_OH_PSR}). 
Currently, no study has ever tried to search for CH absorption or emission towards pulsars. 
Astrochemically, both OH and CH are simple hydrides expected to be abundant in regions of intermediate extinction, where molecular hydrogen has formed but there is insufficient shielding for CO to reach stable abundance.  
Several studies have demonstrated that OH and CH both trace the so-called "dark molecular gas" (DMG) effectively and correlate well with each other \citep{Xu_2016_OH,Xu_2016_CH,2018ApJS..235....1L,Ningyu_2021_CH}. 
Therefore, it is natural to expect stimulated amplification of pulsar radiation through both OH and CH transitions along the same sightlines.

Murriyang, CSIRO's Parkes Radio Telescope, equipped with the ultra-wide-bandwidth low-frequency receiver (UWL), enables simultaneous recording of three CH 3.3 GHz transitions \citep{Hobbs_UWL_2020}. 
As \citet{Liu_HI_2021} demonstrated the capability of generating pulsar pulse-phase-resolved HI spectra using the UWL, it is straightforward to extend this technique to search for CH maser emission and to then investigate its variability on millisecond timescales towards PSR J1644$-$4559. 
In this work, we present the first search for pulsar absorption and pulsed CH maser emission stimulated by PSRJ1644$-$4559, achieving high sensitivity with the Parkes UWL.

\section{Observations and Data Reduction}
\label{sect:Obs} 

Observations of the hyperfine structure lines between the $\Lambda$-doublet of CH in the $^{2}\Pi_{1/2}, J=1/2$ state towards PSR~J1644$-$4559 were conducted from February 2022 to April 2023 using the UWL on Murriyang, CSIRO’s Parkes Radio Telescope. 
The total integration time for each of the three $\Lambda$-doubling transitions of CH in the $^{2}\Pi_{1/2}, J=1/2$ state, at the rest frequencies of 3335.481 (main line $F=1-1$), 3263.794 (lower satellite line $F=0-1$), and 3349.193 MHz (upper satellite line $F=1-0$), is $\sim$71 hours. 
Similar to \citet{Liu_HI_2021}, the data were recorded in baseband mode across the entire subband 20 with frequency coverage from 3264--3392 MHz,  allowing simultaneous detection of the three CH transitions at 3264, 3335, and 3349 MHz. 
The velocity at the 3264 MHz band edge, relative to the rest frequency of 3263.794 MHz, is $\sim -18.9$ \kms. 
Considering the projected velocity of the telescope with respect to the local standard of rest (LSR), it is still capable of detecting spectral components with LSR velocities more negative than $-30$ \kms. 
Since the OH absorption and emission observed towards PSRJ1644$-4559$ are concentrated at velocities below $-30$ \kms, we are able to detect these clouds in CH 3263.794 MHz transition at subband 20.
A calibrator source (CAL) was injected while pointing $\sim$1 degree away from the pulsar position (RA=16$^{\rm h} $44$^{\rm m}$49.2$^{\rm s}$, Dec= $-45$\dg 59$'$ 09.7$''$) after each hour of integration to perform flux calibration.

The pulsar baseband data were coherently dedispersed and folded synchronously with the pulsar period (455.078 ms) for each 10-second integration using the \texttt{DSPSR} package \citep{Willem_dspsr_2011}. This produced datasets with 128 phase bins and 262144 frequency channels, achieving a frequency resolution of 0.49 kHz (or $\sim$0.044 \kms\ velocity resolution) and a time resolution of 78.1 ms. 
The CAL baseband data were folded in sync with the CAL period ($\sim$0.09 s) for each 10-second integration, resulting in data cubes with 32 phase bins and a frequency resolution of 0.49 kHz. 
Flux calibration was performed using the \texttt{pac} command from the \texttt{PSRCHIVE} package \citep{Willem_psrchive_2012}, applying calibration solutions from the observatory based on observations of flux calibrators 0407$-$658 and 1934$-$638 taken nearest in time to our observations.
The final calibrated data are in units of Jy.

The dataset is potentially contaminated by radio frequency interference (RFI). 
For strong, wide-band RFI, the affected sub-integrations were identified and subsequently removed from the dataset. 
For narrow-bandwidth, time-variable RFI,  
channels with values exceeding $4\sigma_{\rm off}$, where $\sigma_{\rm off}$ is the standard deviation of each frequency channel over a 10-minute pulsar-off observation, were removed from subsequent integration.

Assuming a source that  fills the telescope beam, the brightness temperature $T_{\rm b}$ can be derived from the flux density $S_{\nu}$ observed at a frequency $\nu$ using the Rayleigh-Jeans relation 
\begin{equation}
T_{\rm b}= \frac{c^2 S_{\nu}}{2\nu^2 k_{\rm B}\Omega},
\label{e:tb_S_parkes}
\end{equation}
\noindent where $k_{\rm B}$ is the Boltzmann constant, $c$ is the speed of light, and $\Omega$ the beam solid angle. 
Based on the Gaussian main beam width of 0.13 degrees at subband 20 for the Parkes UWL \citep{Hobbs_UWL_2020}, the theoretical conversion factors for three CH $\Lambda$-doubling spectral lines at 3264, 3335, and 3349 MHz are (0.594, 0.569, and 0.564) K / Jy, respectively.

\section{Results and discussion}
\label{sect:analysis}

The radiative transfer equations for pulsar-ON and -OFF measurements, after subtraction of terms that are constant across the band, are given by
\begin{align}
 &T_\mathrm{\rm b}^\mathrm{ON} (v) =(T_\mathrm{ex}-T_\mathrm{bg})(1-e^{-\tau'(v)})-T_\mathrm{psr}(1-e^{-\tau(v)}),\\
 &T_\mathrm{\rm b}^\mathrm{OFF} (v)=(T_\mathrm{ex}-T_\mathrm{bg})(1-e^{-\tau'(v)}),
\label{eq:rt_OH}
\end{align}
where $T_\mathrm{ex}$ and $T_\mathrm{psr}$ denote the CH excitation temperature and pulsar temperature, respectively. 
The term $\tau'(v)$  represents the optical depth profile resulting from CH at any distance within the Parkes beam, while $\tau(v)$ originates from CH within the pencil beam extending only from the pulsar to the observer.\footnote{It is implicitly assumed in this formulation that both the continuum background and the CH fill the beam.}  
The background (i.e, beyond the CH) brightness temperature, $T_\mathrm{bg}$, consists of contributions from the isotropic CMB (2.73 K) and from that portion of the Galactic continuum emission originating beyond the CH. 
However, it is not possible to determine observationally how much of the Galactic continuum lies behind the CH emission. 
Therefore, the lower limit of the background temperature $T_\mathrm{bg}^{\rm lo}$ is 2.73 K, while the upper limit $T_\mathrm{bg}^{\rm up}$ includes contributions from both the CMB and the Galactic continuum. 
We followed the method outlined in \cite{SPLASH_2022} to estimate an upper limit for the continuum background temperature $T_\mathrm{bg}^{\rm up}$ at the pulsar position by scaling the S-PASS survey data at
2300 MHz \citep{SPASS_2019}. 
$T_\mathrm{bg}^{\rm up}$ can be expressed as 
\begin{equation}
T{\rm_{bg}^{\rm up}} = 2.73 + T{\rm_{bg2300}}(\nu{\rm_{CH}}/2300\ \rm MHz)^{\beta}, \\
\label{eq:tbg_CMB_GS} 
\end{equation}
where $T_{\rm bg2300}\approx7$ K is obtained from S-PASS survey data at the source position, and $\beta\approx$ -2.1 is the spectral index at the source position as estimated from the S-PASS map at 2300 MHz and the CHIPASS map at 1395 MHz \citep{CHIPASS_2014}. 
The corresponding $T{\rm_{bg}}^{\rm up}$ at 3.3 GHz is $\sim$6.0 K.

The baselines for the optical depth spectra were generally flat and were fitted and subsequently subtracted using a second-order polynomial. 
The pulsar-ON and -OFF spectra were significantly affected by instrumental baselines, making it challenging to unambiguously fit and remove the entire baseline beneath the spectral lines. 
To address this, the spectra were divided into three separate velocity segments containing the spectral components, and their baselines were each fitted and subtracted using a second-order polynomial, in preparation for Gaussian decomposition, as described in Section~\ref{sect:gd}.

\begin{figure*}
\centering
\includegraphics[width=0.45\linewidth]{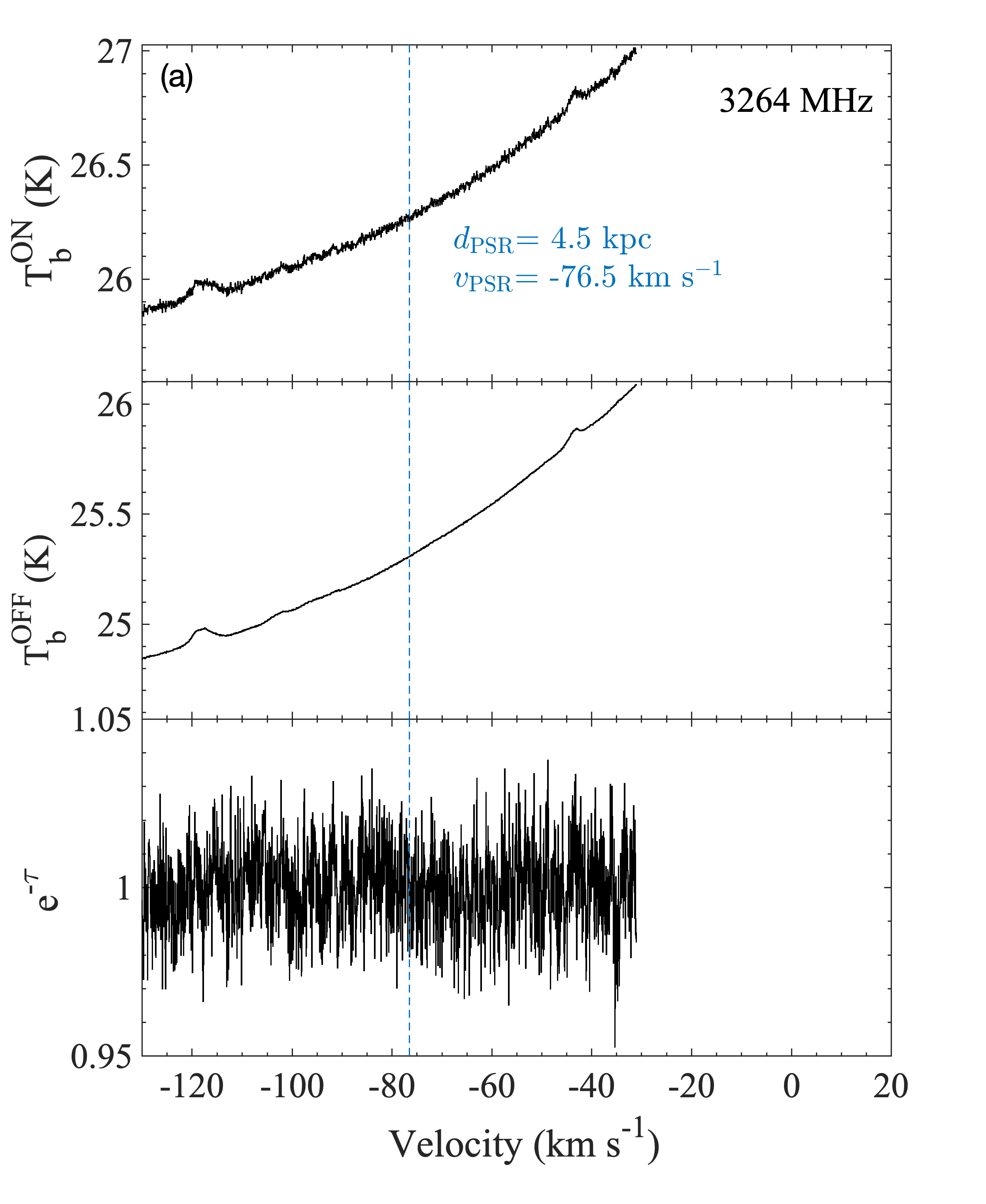}
\includegraphics[width=0.45\linewidth]{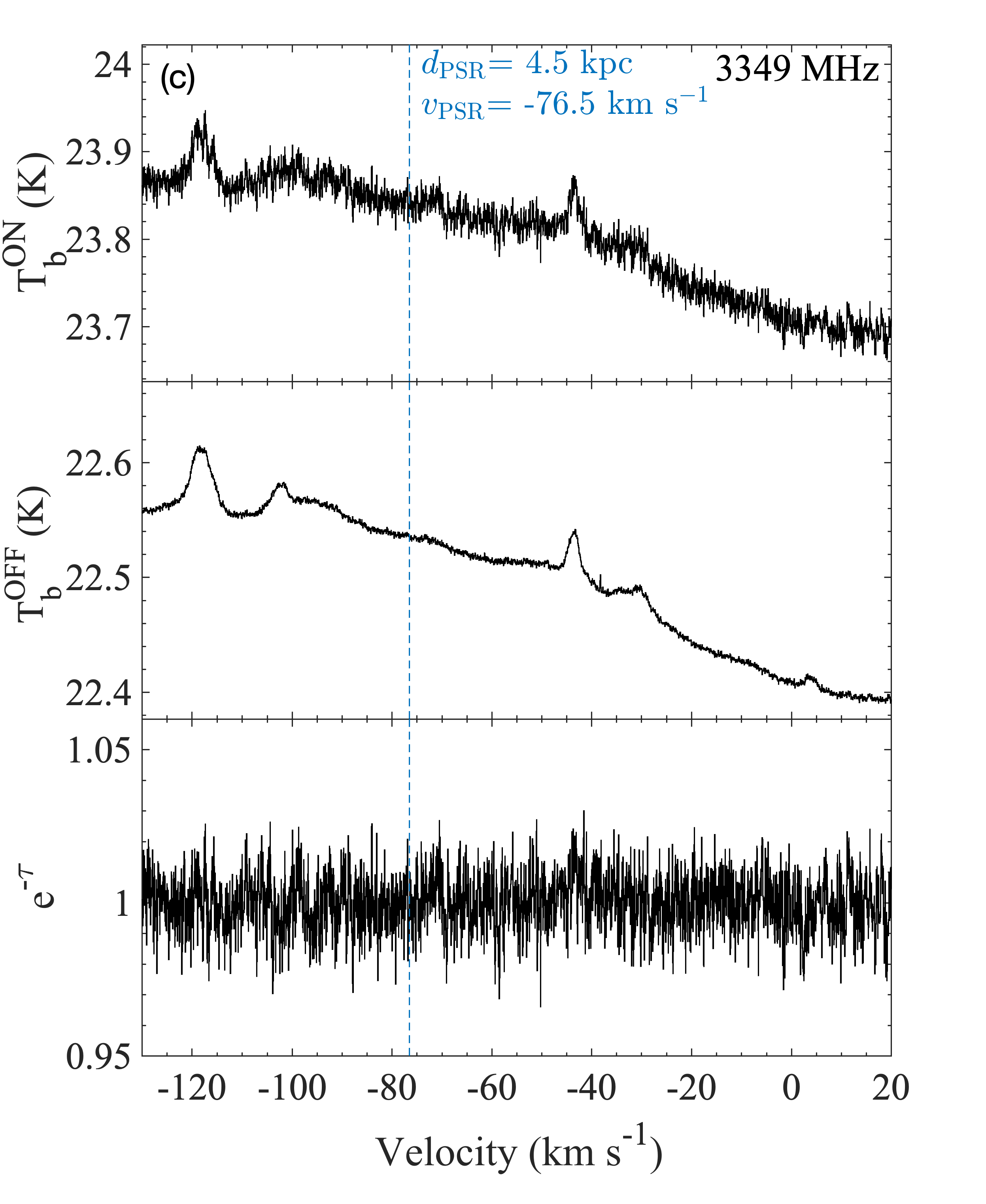}
\includegraphics[width=0.45\linewidth]{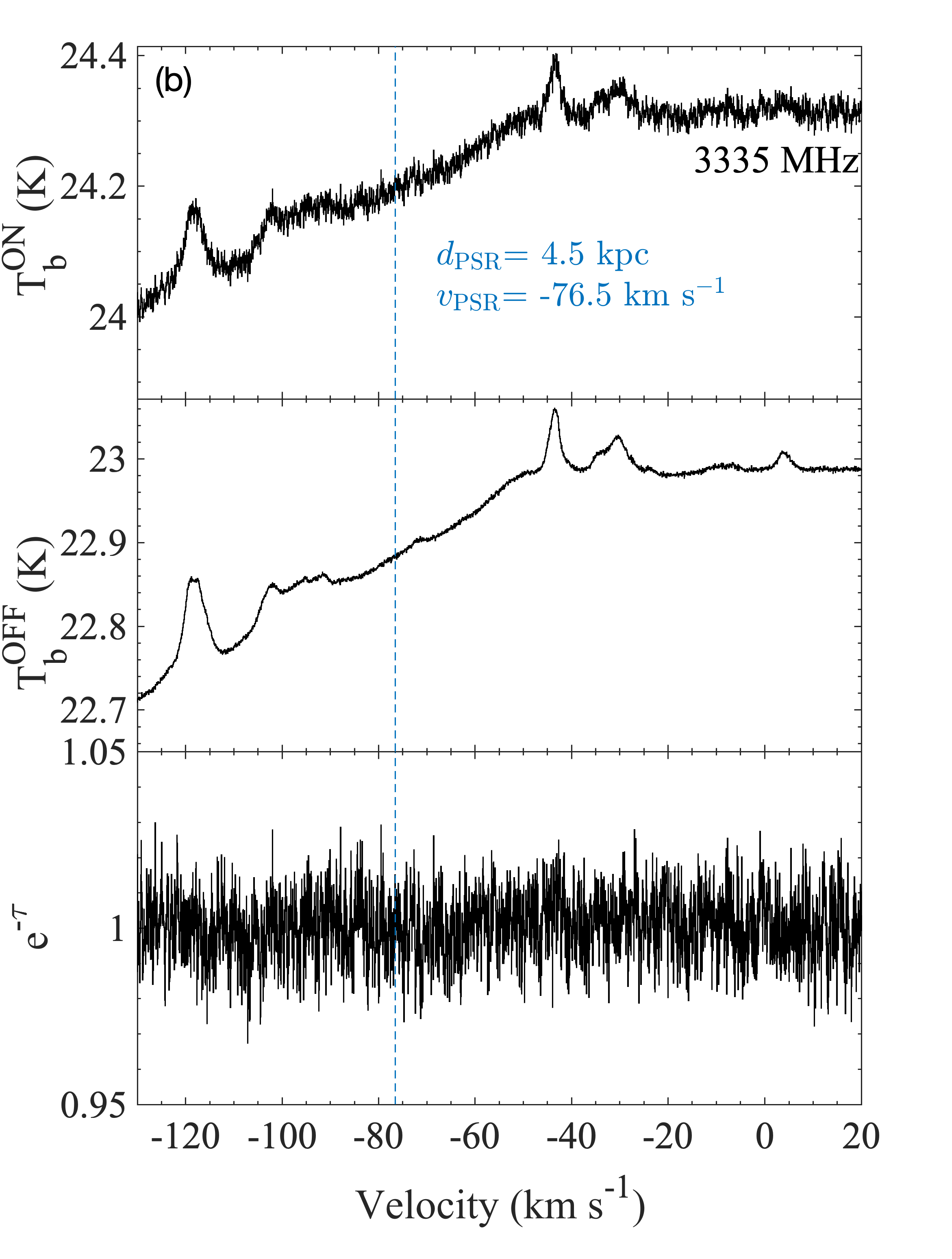}
\caption{ (a) Top and middle panels show the raw pulsar-ON and -OFF spectra observed at 3264 MHz in brightness temperature. The bottom panel displays $e^{-\tau}$, the difference between the pulsar-ON and -OFF spectra scaled by the mean value of the baseline of this difference spectrum, after baseline removal. The blue dashed line indicates the LSR velocity of the pulsar, based on its kinematic distance of 4.5 kpc. (b) Same as (a), observed at 3349 MHz. (c) Same as (a), observed at 3335 MHz.
}
\label{fig:CH_emi}
\end{figure*}

Figure~\ref{fig:CH_emi} presents the first spectra of the three CH hyperfine transitions observed toward a pulsar (PSR~J1644$-$4559). 
The spectra, collected during pulsar-ON and pulsar-OFF phases without baseline removal, are shown along with the corresponding optical depth spectra.
Three main CH emission components are observed within the Parkes beam at velocities of $\sim-110$, $-45$, and 5 \kms\ in both pulsar-ON and pulsar-OFF spectra. 

The $-110$ and $-45$ \kms\ features align well with the OH absorption and emission detected toward this pulsar during OFF periods by \citet{Weisberg_etal_2005}. 
The same LSR velocities of CH and OH spectral lines indicate the same kinematic distances of these two clouds, suggesting a close spatial association between CH and OH molecules. 
Such a relationship between the two species has been observed extensively from quiescent dark clouds to HII regions (e.g. \citealt{1976ApJS...31..333R,Suutarinen_TMC1_CH_2011,Xu_2016_CH,Ningyu_2021_CH,2022ApJ...930..141J}).
The OH features at $-110$ and $-45$ \kms\ exhibit both emission and absorption, demonstrating departures from local thermodynamic equilibrium (LTE). 
Given the similar level structure of the CH radical and OH, non-LTE conditions are also expected for CH transitions. 
The weakest CH component detected at 5 \kms\ was not detected in the four OH $\Lambda$ doublet transitions in $^{2}\Pi_{3/2}, J=3/2$ state(1612, 1665, 1667, and 1720 MHz), likely due to the short integration time of $\sim$5 hours in \citet{Weisberg_etal_2005}.

The kinematic distance of PSR~J1644$-$4559 is $\sim$4.5 kpc \citep{Verbiest_PSR_dist_2012}, corresponding to the velocity with respect to Local Standard of Rest (LSR) of $v_{\rm PSR} = -76.5$ \kms. 
Any features with velocities more negative than $v_{\rm PSR}$ are more distant than the pulsar, and thus no features are expected in the optical depth spectra for the CH component at $\sim-110$ \kms.  
For the $-45$ and $5$ \kms\ features, which are in front of the pulsar, no significant signal is detected in the optical depth spectrum, $\tau(v)$.

\subsection{Gaussian Decomposition and Column Density Estimation}
\label{sect:gd}

Although no significant signals were detected in the optical depth spectra, clear emission components are present in both the pulsar-ON and -OFF spectra. 
These components can still be analyzed to constrain the properties of CH clouds within the entire Parkes beam. 
Additionally, upper limits can be estimated on the column densities of CH clouds pierced by the pulsar beam.

\begin{figure*}
\centering
\includegraphics[width=0.496\linewidth]{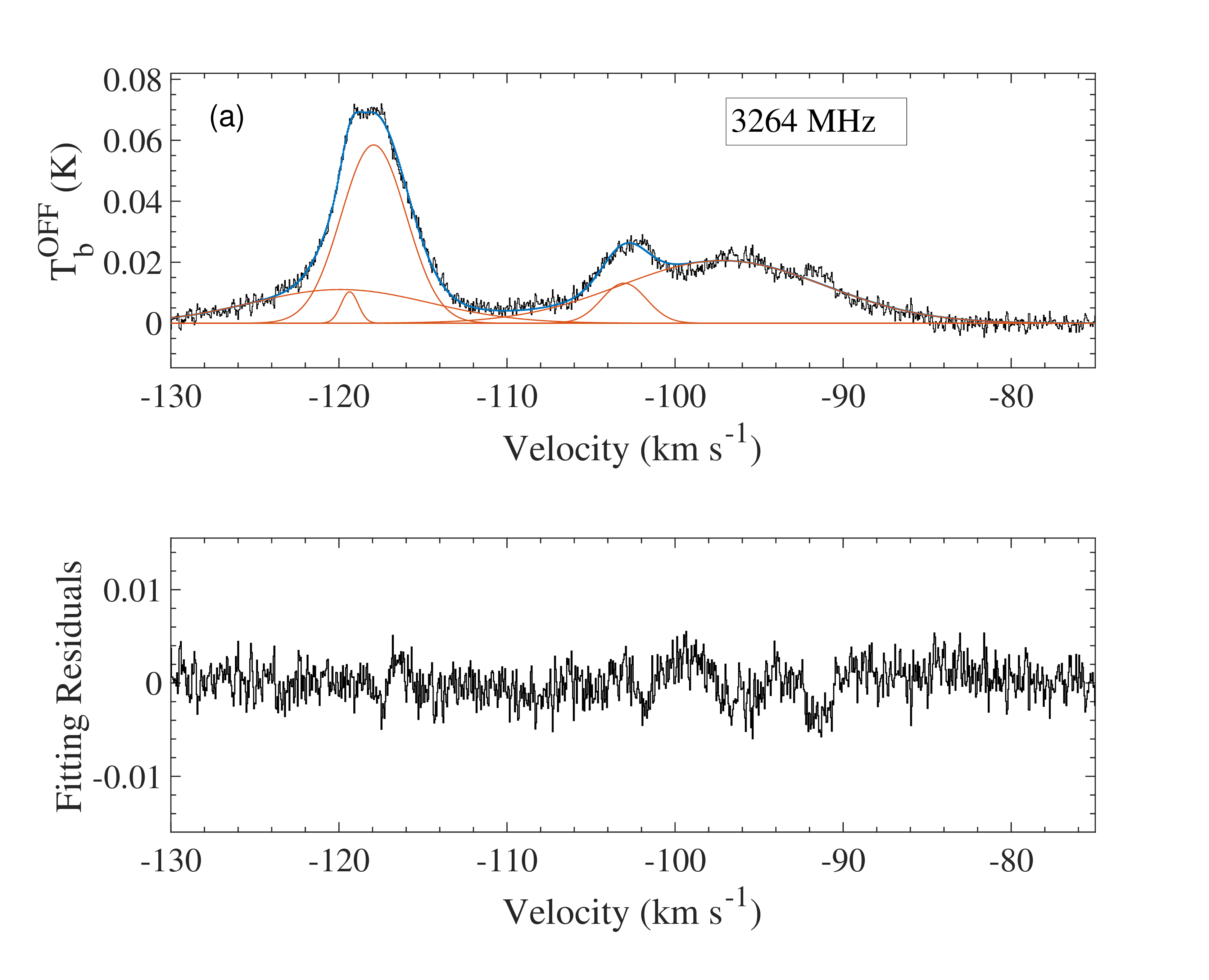}
\includegraphics[width=0.496\linewidth]{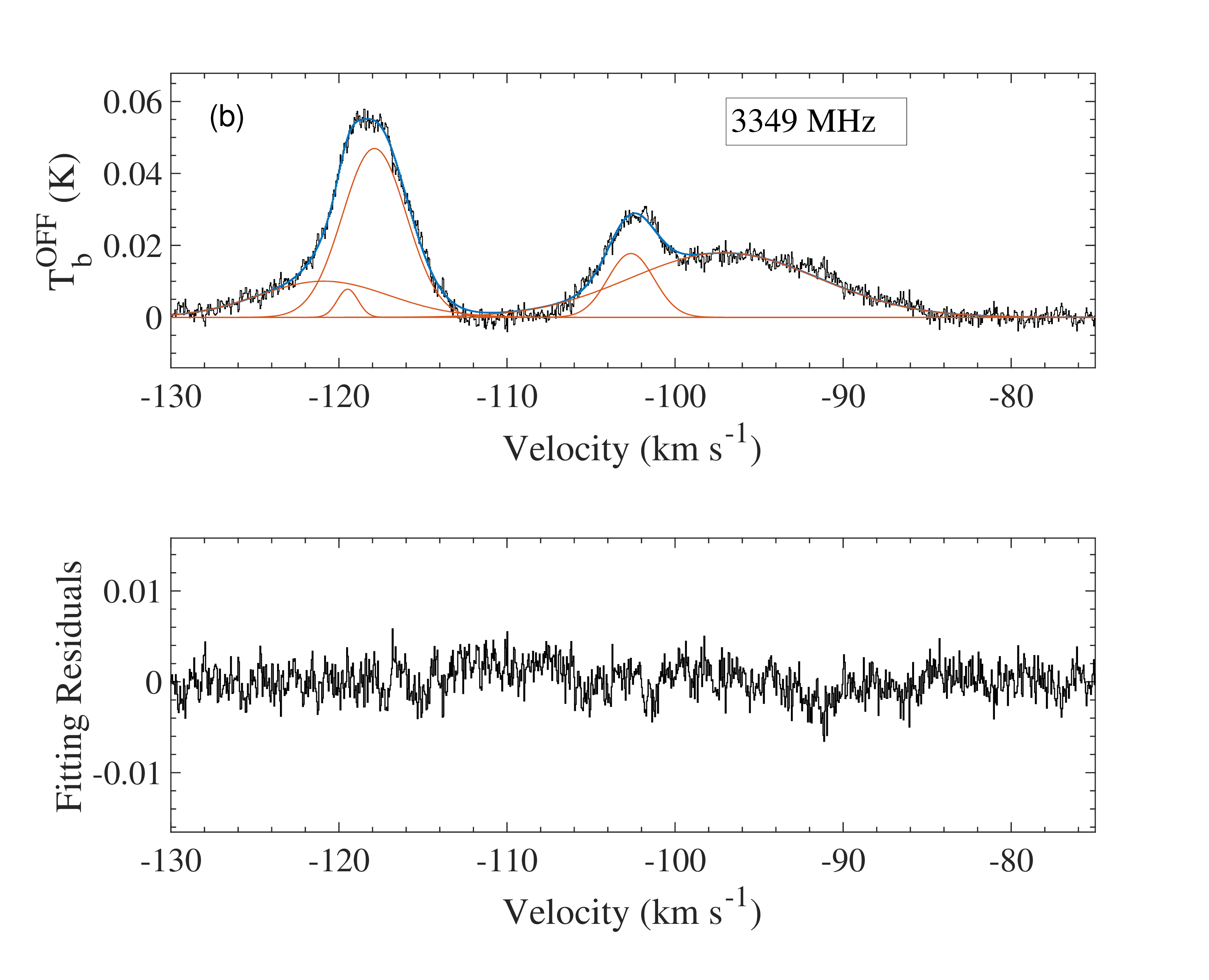}
\includegraphics[width=0.496\linewidth]{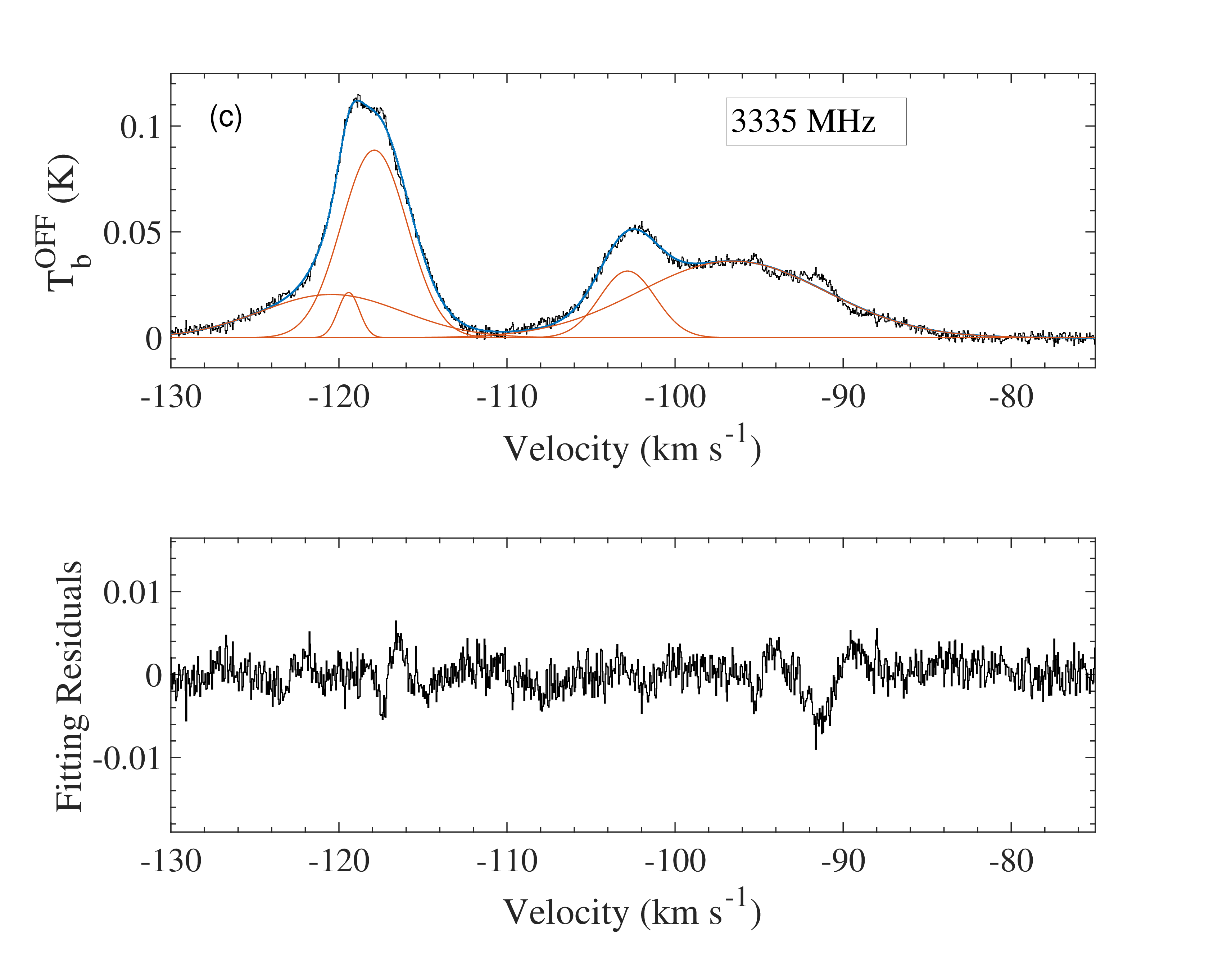}
\caption{(a) For the 3264 MHz transition. Top panel: The black line shows the CH emission during the pulsar-OFF period at $\sim110$ \kms. The red dashed lines represent the five Gaussian components, and the blue line illustrates the final fitted CH emission.  Bottom panel: The residuals from the fit.
(b) Same as (a) for the 3349 MHz transition.
(c) Same as (a) for the 3335 MHz transition. 
}
\label{fig:-100_fit}
\end{figure*}

\begin{figure*}
\centering
\includegraphics[width=0.496\linewidth]{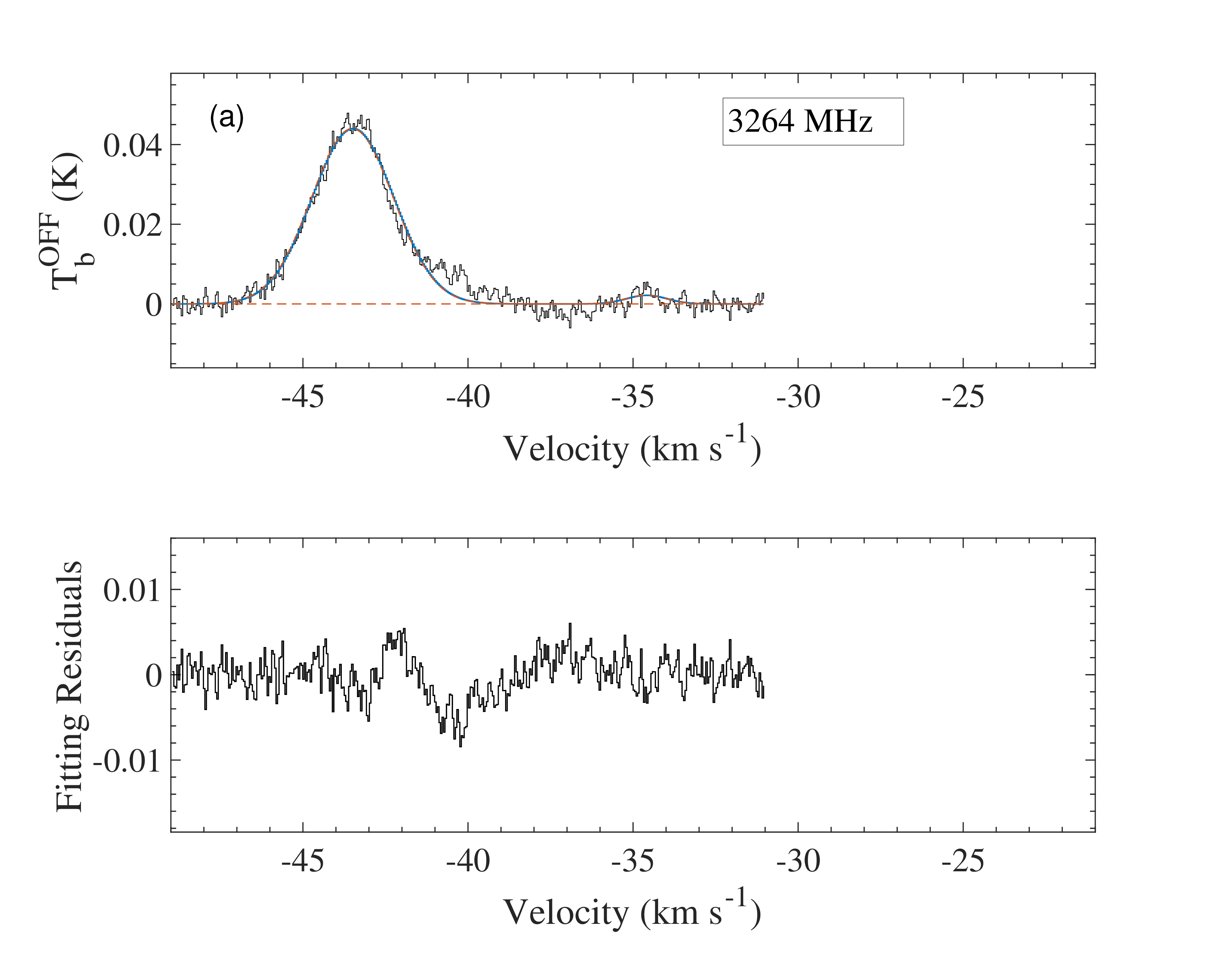}
\includegraphics[width=0.496\linewidth]{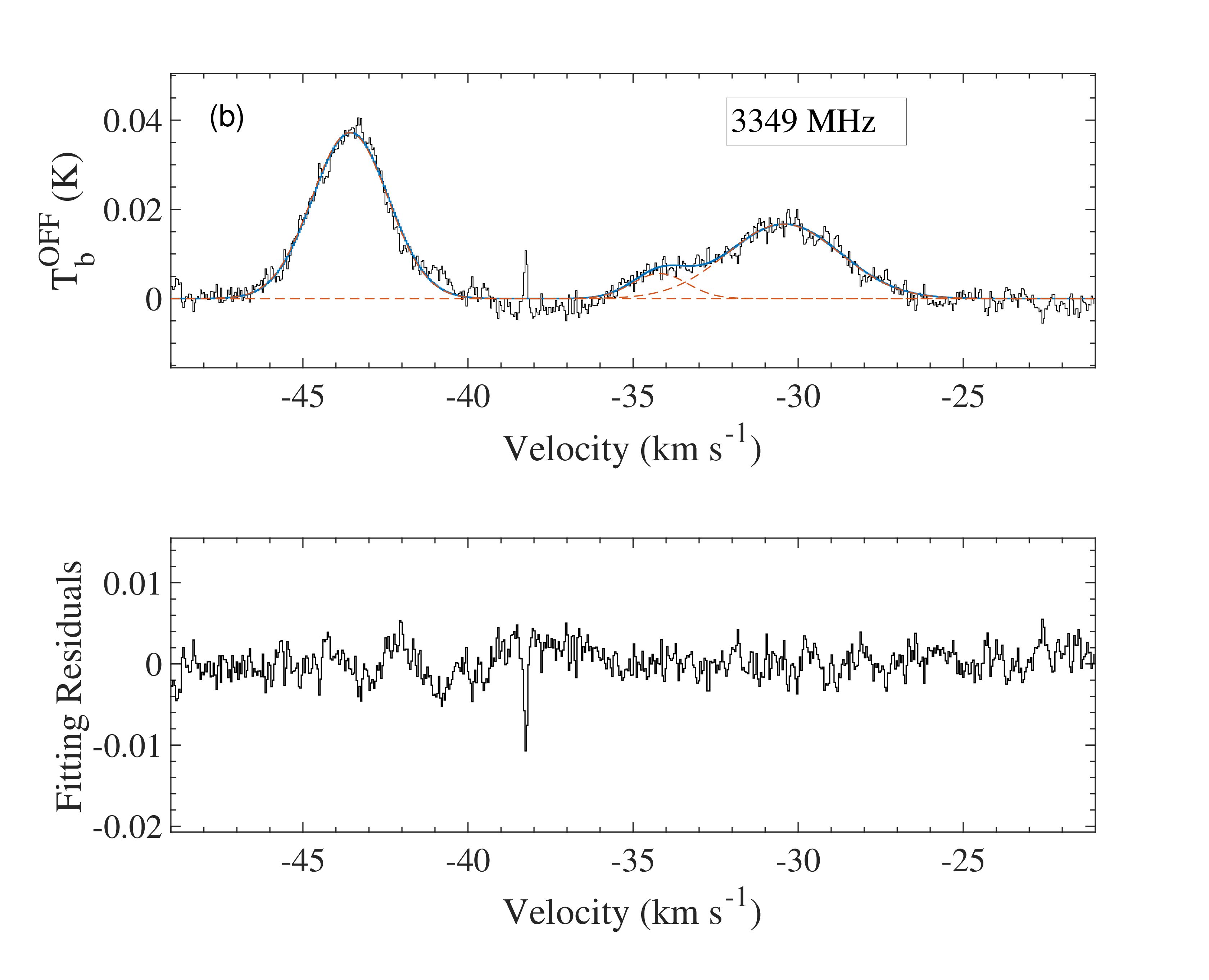}
\includegraphics[width=0.496\linewidth]{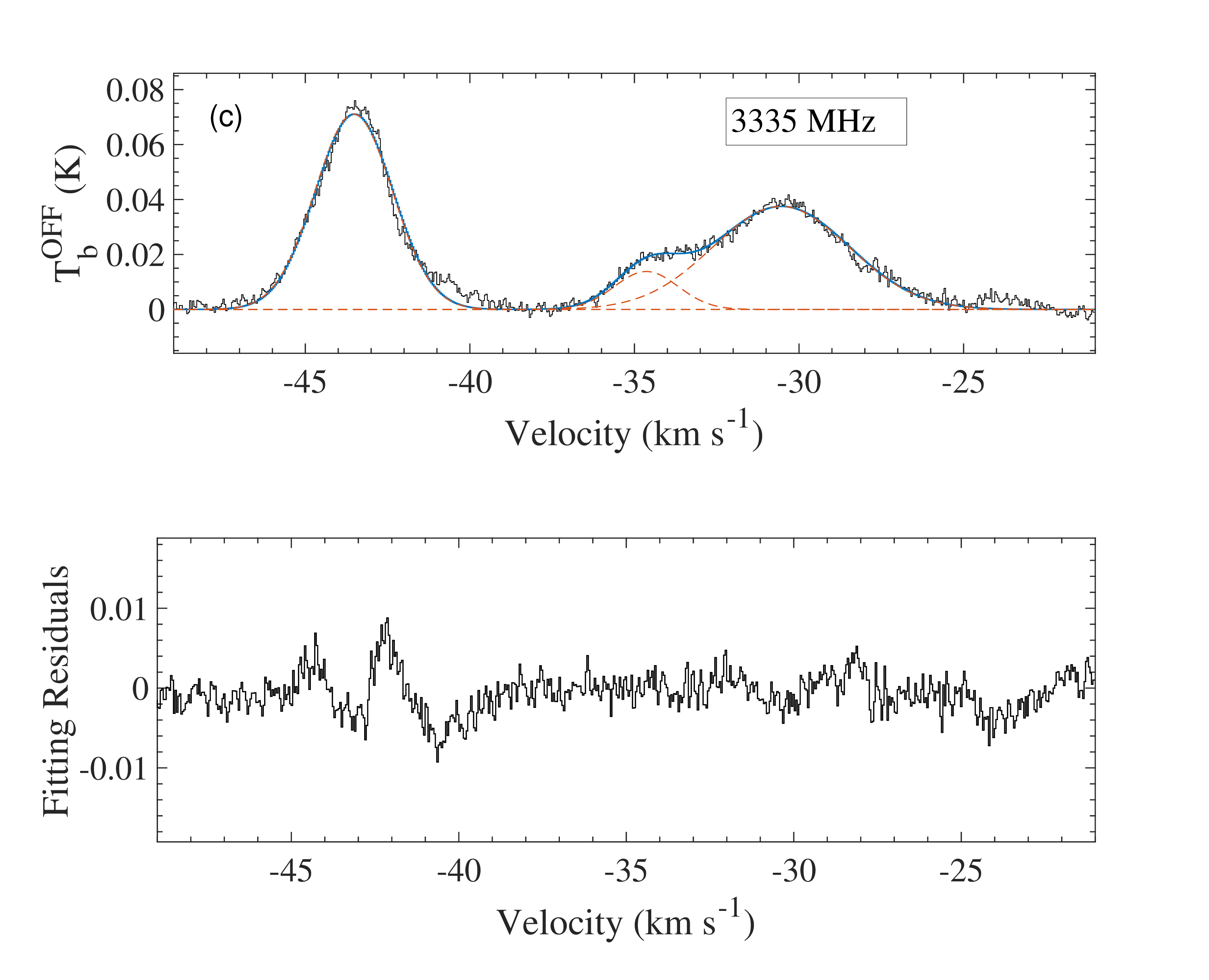}
\caption{(a) For the 3264 MHz transition. Top panel: The black line shows the CH emission during the pulsar-OFF period at $\sim-45$ \kms. The spectrum is truncated at $\sim-30$ \kms\ due to the band edge. The red dashed lines indicate the two Gaussian components, and the blue line illustrates the final fitted CH emission.  Bottom panel: The residuals from the fit.
(b) For the 3349 MHz transition. Top panel: The black line shows the CH emission during the pulsar-OFF period at $\sim-45$ \kms. The red dashed lines represent the three Gaussian components. 
(c) Same as (b) for the 3335 MHz transition. 
}
\label{fig:-40_fit}
\end{figure*}

\begin{figure*}
\centering
\includegraphics[width=0.496\linewidth]{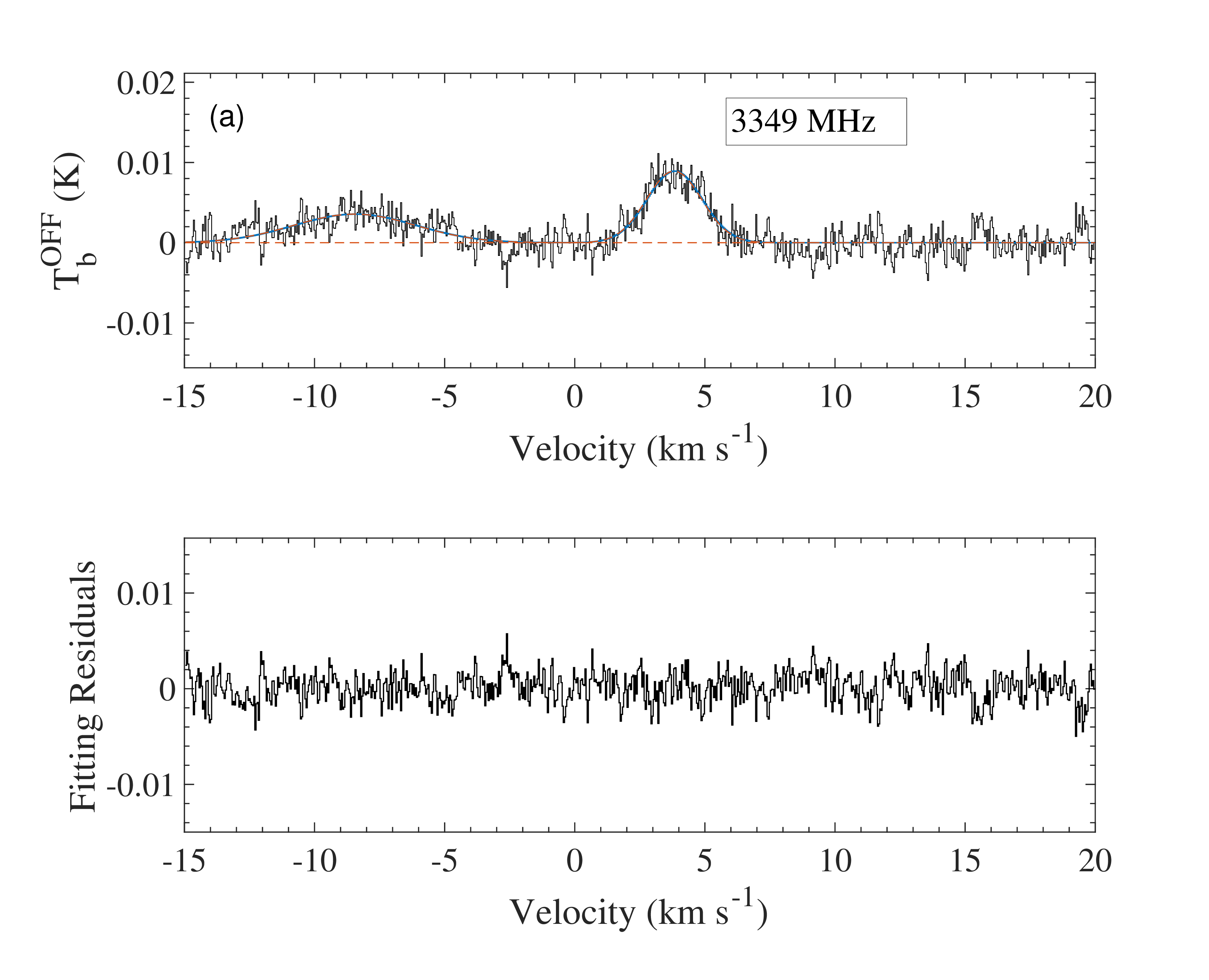}
\includegraphics[width=0.496\linewidth]{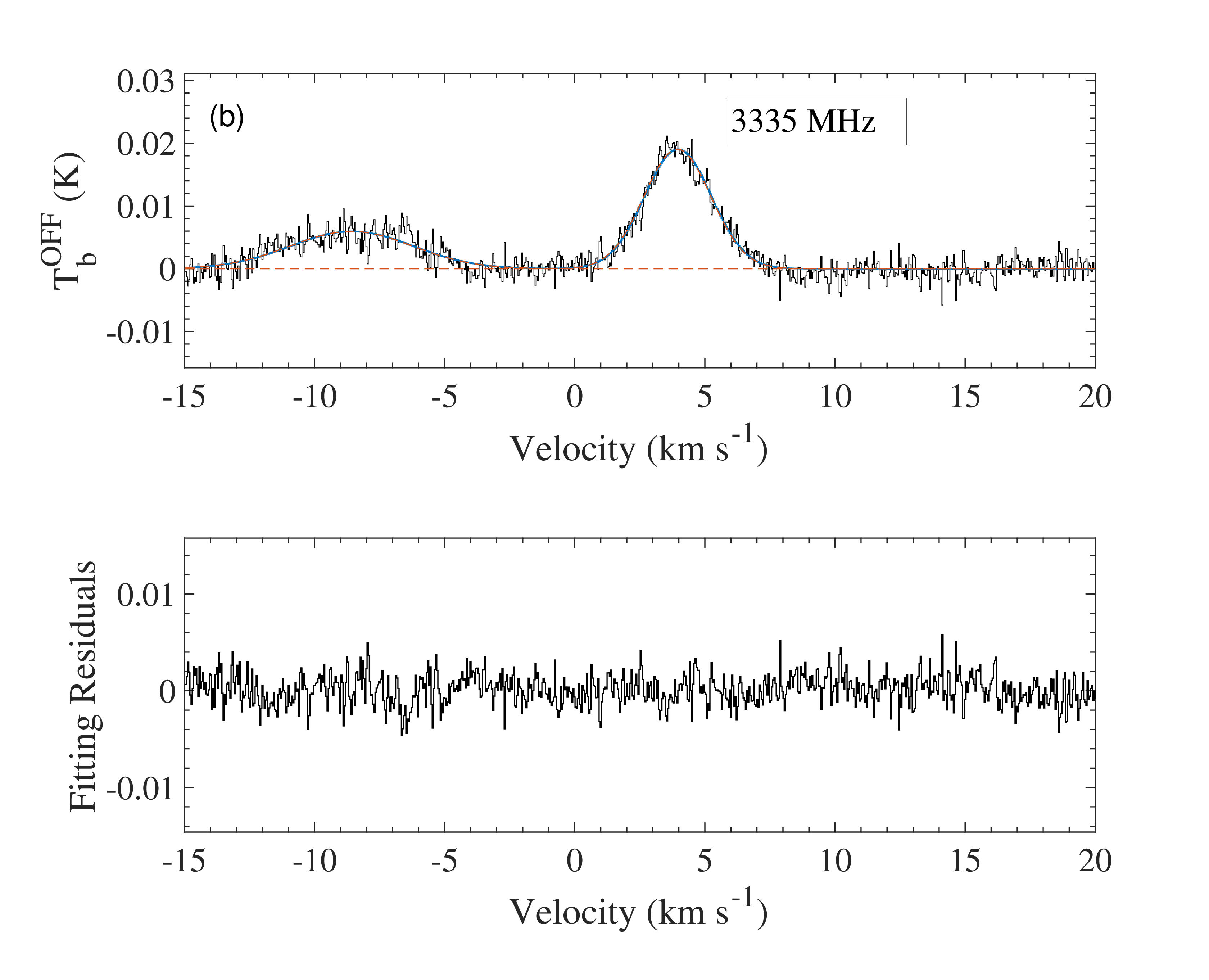}
\caption{(a) For the 3349 MHz transition. Top panel: The black line shows the CH emission during the pulsar-OFF period at $\sim5$ \kms. The red dashed lines indicate the two Gaussian components, and the blue line illustrates the final fitted CH emission.  Bottom panel: The residuals from the fit.
(b) Same as (a) for the 3335 MHz transition. No data were recorded near $\sim5$ \kms\ for the 3264 MHz transition due to the band-edge frequency cutoff. 
}
\label{fig:10_fit}
\end{figure*}

Due to the short duty cycle ($\sim$1.8\%) of PSR~J1644$-$4559, the pulsar-ON spectra have a significantly higher noise level than the pulsar-OFF spectra. 
To obtain reliable Gaussian decomposition, we first use the curve fitting tool \texttt{cftool} in Matlab to estimate initial guesses for the Gaussian components from the pulsar-OFF spectra $T_{\rm b}^{\rm OFF}$, which can be expressed as 
\begin{equation}
T_{\rm b}^{\rm OFF}(v) = \sum_{n=0}^{N-1} T_{\text{c},n} \cdot e^{- 4 \ln{2} \left(v-v_{0,n}\right)^2/ \Delta v_{0,n}^2},
\label{e:tau_abs}
\end{equation}
\noindent where $T_{\text{c},n}$, $v_{0,n}$, $\Delta v_{0,n}$ are the amplitude, central velocity, and FWHM of the $n$th Gaussian feature, respectively.
The final fitting results were refined using \texttt{lsqcurvefit} with a least-squares approach, allowing amplitudes, central velocities, and line widths to vary within 20\%  with respect to the initial estimates. 
The Gaussian fitting results for the three CH transitions at $\sim -110$, $-45$, and $5$ \kms\ are shown in Figure~\ref{fig:-100_fit} to Figure~\ref{fig:10_fit}.
The derived parameters are listed in Table~\ref{tab:fit_Gauss}. 
A total of ten Gaussian components were identified along the line of sight. 
Due to the frequency cutoff at 3264 MHz in subband 20, spectra at velocities greater than $-30$ \kms\ can not be recorded for the 3264 MHz transition.

Under the assumption of small optical depth, the column density for the CH 3335 MHz transition detected across the entire Parkes beam in the pulsar-OFF spectra can be calculated  using \citep{1979A&A....73..253G}:
\begin{equation} 
N(\mathrm{CH})_{\rm OFF} = 2.82\times10^{14}   T_{\mathrm{ex}}  \frac{T_{c,n}  \Delta v_{0,n}}{T_{\mathrm{ex}} - T_{\mathrm{bg}}}\delta, 
\label{eq:ch_off}
\end{equation}
\noindent where $T_{c,n}$ and $\Delta v_{0,n}$ are the peak brightness temperature and FWHM of the fitted Gaussian components, as listed in Table~\ref{tab:fit_Gauss}. 
The statistical weights of the levels $\delta$= 2 for 3264 and 3349 MHz, and $\delta$= 1 for the 3335 MHz transition line\citep{1976ApJS...31..333R}. 
Since no significant $\tau(v)$ signal is detected in the pulsar ON$-$OFF spectra, we cannot simultaneously determine $\tau(v)$ and $T_{\rm ex}$. 
\citet{Ningyu_2021_CH} directly estimated the excitation temperatures for the CH transitions at 3335, 3264, and 3349 MHz using ON/OFF observations of extragalactic continuum sources.
Their Monte Carlo analysis determined that the excitation temperatures for the 3335 MHz main line and the 3264 and 3349 MHz satellite lines are concentrated within the ranges $[-5, 0]$, $[-5, 0]$, and $[-10, -5]$ K, respectively. 
Based on these results, we adopted the average values of $-2.5$, $-2.5$, and $-7.5$ K for $T_{\rm ex}$ at the 3264, 3335, and 3349 MHz transitions, respectively. 
We estimated the lower and upper limits of the CH column densities for the three transitions by incorporating the lower and upper bounds of the background temperature, with $T_\mathrm{bg}^{\rm lo} = 2.73$ K and $T_\mathrm{bg}^{\rm up} = 6$ K. 
The resulting column densities for all ten CH components at 3335 and 3349 MHz, as well as seven features at 3264 MHz, are listed in Table~\ref{tab:ch_N_off}.

The calculated column densities across three CH transitions from different Gaussian components during pulsar-OFF range from $0.2$ to $9.8 \times 10^{13} , \rm cm^{-2}$, excluding feature 8 at the 3264 MHz transition due to its extremely low intensity.
Previous chemical models have demonstrated a close correlation between CH and $\rm H_2$, as CH forms primarily through reactions with $\rm H_2$\citep{1973ApL....15...79B,1975ApJ...199..633B}.
Observationally, the relationship between CH and $\rm H_2$ column densities has been established in various studies (e.g., \citep{2002A&A...391..693L,2008ApJ...687.1075S,Ningyu_2021_CH}).
We adopted an average CH abundance relative to $\rm H_2$ of $3.5 \times 10^{-8}$ from \citet{Ningyu_2021_CH}, derived from observations of translucent clouds along various sightlines. 
Using this abundance, the $\rm H_2$ column densities for the ten clouds within the Parkes beam range from $0.6 \times 10^{20}$ to $2.8 \times 10^{21} , \rm cm^{-2}$, suggesting they are likely diffuse and translucent clouds \citep{2011piim.book.....D}.

\begin{figure*}
\centering
\includegraphics[width=0.8\linewidth]{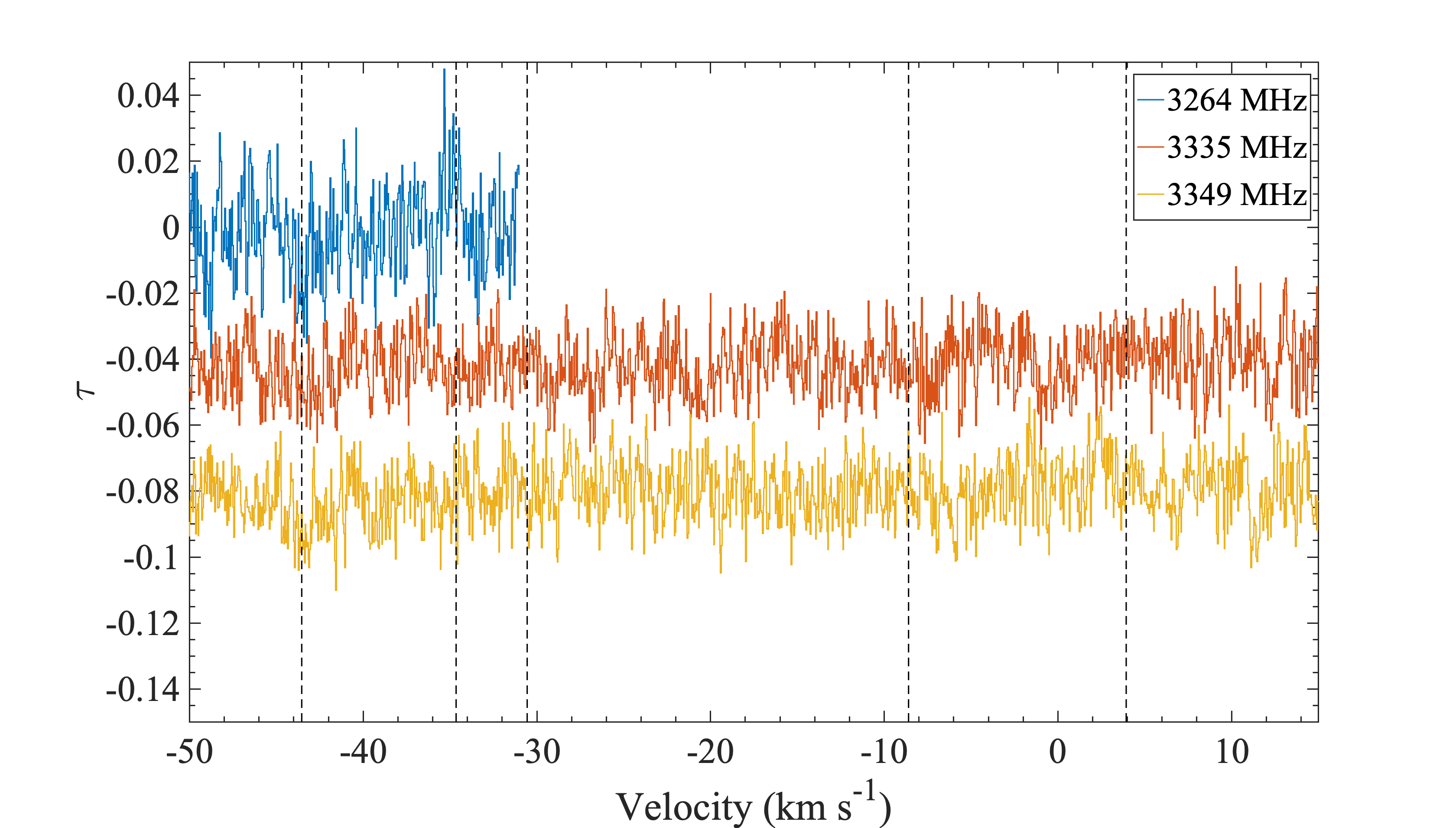}
\caption{Optical depth spectra, $\tau$= $1 - \exp(-\tau)$, of the CH $\Lambda$-doublet transitions towards PSR~J1644$-$4559. The blue, red, and yellow lines represent the transitions at 3264 MHz, 3335 MHz (offset by \(-0.04\) in $\tau$, and 3349 MHz (offset by \(-0.08\) in $\tau$, respectively. The dashed lines indicate the central velocities of five Gaussian CH emission components detected in the pulsar-OFF spectra.
}
\label{fig:ch_tau_all}
\end{figure*}

For CH traced by the pulsar pencil beam, we consider only the components located in the foreground of the pulsar. 
The optical depth spectra of three CH transitions in front of the pulsar are shown in Figure~\ref{fig:ch_tau_all}. 
Five emission features are detected in the pulsar-OFF spectra within these velocity ranges, with peak positions marked by dashed lines. 
However, no clear absorption or emission features are observed in the optical depth spectra. 
We therefore estimate an upper limit for the CH column density, $N(\mathrm{CH})_{\rm rms}$, by integrating the $\tau$ spectra over the same FWHM velocity range as each Gaussian component obtained from the pulsar-OFF spectra, expressed as
\begin{equation}
N(\rm CH)_{\rm rms} = 2.82\times 10^{14} cm^{-2} | T_{ex} \int_{v_{0,n}-\Delta v_{0,n}/2.}^{v_{0,n}+\Delta v_{0,n}/2.} \tau(v) dv|\delta,
\label{eq:Nch_rms}
\end{equation}
\noindent $v_{0,n}$ and $\Delta v_{0,n}$ are listed in Table~\ref{tab:fit_Gauss}. 
The corresponding upper limits on the column density are provided in Table~\ref{tab:ch_N_off}.

The pulsar beam ($\sim$0.03", calculated from pulsar pulse scattering; \citep{2013MNRAS.434...69L}) is much narrower than the FWHM of the Parkes beam at subband 20 ($\sim$7.8 arcmin; \citep{Hobbs_UWL_2020}). 
Comparing the column density upper limits within the pulsar beam to the column densities observed in the Parkes beam provides insight into the clumpiness and substructure of CH clouds. 
While the non-detection of signal in the $\tau$ spectra makes it difficult to draw definitive conclusions, in some cases (e.g. Feature No. 6, 7, 9 at 3335 MHz and No. 7 at 3349 MHz as listed in Table~\ref{tab:ch_N_off}), the column density upper limits observed within the pulsar beam are significantly lower than the lower limits observed in the Parkes beam.
This suggests that much of the CH structures are not pierced by the pulsar pencil beam, indicating that CH clouds may exhibit clumpiness and substructure in certain regions.

\subsection{Upper Limits of Potential Pulsed CH Maser Amplification Gain}

\citet{Weisberg_etal_2005} discovered pulsed OH maser emission at 1720 MHz with a velocity of $-45$ \kms, stimulated by the photons from PSR~J1644$-$4559, with a shortest maser variability time scale of 14 ms. 
The peak optical depth of the pulsed OH maser line is $\tau_{\rm maser} \sim-0.05$, indicating an unsaturated state with an amplification gain of $e^{-\tau_{\rm maser}}$= 1.05. 
Stimulated emission occurs when the relevant level populations are inverted or pumped by some process, with suitable photons available to stimulate the emission from the overpopulated level \citep{1992ASSL..170.....E,Weisberg_etal_2005}. 
In the case of pulsed OH maser emission towards PSR~J1644$-$4559, the OH transition levels are already inverted in the clouds due to local radiative or collisional processes, while the stimulating photons are generated by the pulsar \citep{Weisberg_etal_2005}. 
The CH radio emission often involves weak maser effects with an energy level structure similar to OH \citep{1992ASSL..170.....E}. 
We thus estimated the upper limits of amplification gain for the potential pulsed CH maser emission by PSR~J1644$-$4559, using $e^{-\sigma_{\tau}}$, where $\sigma_{\tau}$ is the negative value of the standard deviation of $\tau(v)$ spectra for three transitions. 
The standard deviations of the $\tau(v)$ spectra, $\tau_{\rm rms}$, at the 3264, 3335, and 3349 MHz transitions were calculated over the velocity ranges of [$-50$, $-30$] \kms, [$-50$, $10$] \kms, and [$-50$, $10$] \kms, respectively. The calculation follows $\tau_{\rm rms}= \sqrt{\sum^{n}_{i=1}(\tau_{\rm i}-\Bar{\tau})^2/n}$, where $\tau_{\rm i}$ is the optical depth in the $i$th channel, $\Bar{\tau}$ is the averaged optical depth within the velocity range, and $n$ is the number of channels in the velocity range. 
The resulting values of $\sigma_{\tau}=-\tau_{\rm rms}$ are -0.0137, -0.0091, and -0.0089 for 3264, 3335, and 3349 MHz, respectively.  
The corresponding upper limits for amplification gain are 1.014, 1.009, and 1.009, for the transitions at 3264, 3335, and 3349 MHz, respectively. 
In concrete terms, then, for every 1000 photons provided by the pulsar passing through the cloud, no more than [14, 9, 9] additional photons may be stimulated near those frequencies.

\begin{table*}
\begin{center}
\caption[]{Gaussian Decomposition Results of Three CH Transitions.}
\label{tab:fit_Gauss}
\setlength{\tabcolsep}{1.9pt}
\begin{tabular}{c|ccc|ccc|ccc}
\hline\noalign{\smallskip}
   & 3264 MHz & & &3335 MHz  & & &3349 MHz\\
\hline\noalign{\smallskip}
 Features   &$T_{\rm c}$& $v_0$  &$\Delta v_0$ & $T_{\rm c}$  &$v_0$  & $\Delta v_0$ &$T_{\rm c}$ & $v_0$  & $\Delta v_0$  \\
 (No.)                    & (mK)& (\kms) & (\kms)     & (mK) & (\kms) & (\kms)     & (mK)& (\kms) & (\kms)     \\
\hline\noalign{\smallskip}
1	&	10.3	$\pm$	0.7	&	-119.36	$\pm$	0.03	&	1.2	$\pm$	0.1	&	21.4	$\pm$	0.9	&	-119.42	$\pm$	0.02	&	1.5	$\pm$	0.07	&	7.8	$\pm$	0.9	&	-119.48	$\pm$	0.05	&	1.46	$\pm$	0.17	\\
2	&	58.5	$\pm$	0.8	&	-117.93	$\pm$	0.02	&	4.55	$\pm$	0.05	&	88.6	$\pm$	1.2	&	-117.89	$\pm$	0.02	&	4.61	$\pm$	0.05	&	46.9	$\pm$	1.3	&	-117.89	$\pm$	0.04	&	4.48	$\pm$	0.09	\\
3	&	13.1	$\pm$	0.4	&	-103.03	$\pm$	0.04	&	3.12	$\pm$	0.13	&	31.5	$\pm$	0.5	&	-102.83	$\pm$	0.02	&	3.92	$\pm$	0.07	&	17.7	$\pm$	0.4	&	-102.63	$\pm$	0.03	&	3.24	$\pm$	0.09	\\
4	&	11	$\pm$	0.6	&	-119.9	$\pm$	0.18	&	12.47	$\pm$	0.41	&	20.4	$\pm$	1	&	-120.44	$\pm$	0.14	&	9.95	$\pm$	0.18	&	10	$\pm$	0.9	&	-120.9	$\pm$	0.35	&	9.14	$\pm$	0.33	\\
5	&	20.5	$\pm$	0.2	&	-97.19	$\pm$	0.1	&	14.22	$\pm$	0.18	&	36.2	$\pm$	0.2	&	-96.58	$\pm$	0.07	&	13.02	$\pm$	0.11	&	17.9	$\pm$	0.2	&	-97.07	$\pm$	0.09	&	13.24	$\pm$	0.16	\\
6	&	44	$\pm$	0.4	&	-43.5	$\pm$	0.01	&	2.92	$\pm$	0.03	&	71.1	$\pm$	0.5	&	-43.53	$\pm$	0.01	&	2.77	$\pm$	0.02	&	37.2	$\pm$	0.4	&	-43.58	$\pm$	0.01	&	2.72	$\pm$	0.03	\\
7	&	--	&	--	&	--	&	37.5	$\pm$	0.4	&	-30.55	$\pm$	0.04	&	4.87	$\pm$	0.09	&	16.7	$\pm$	0.3	&	-30.44	$\pm$	0.05	&	4.04	$\pm$	0.12	\\
8	&	2.1	$\pm$	0.7	&	-34.63	$\pm$	0.27	&	1.44	$\pm$	0.66	&	13.8	$\pm$	0.6	&	-34.64	$\pm$	0.05	&	2.28	$\pm$	0.13	&	5.6	$\pm$	0.5	&	-34.2	$\pm$	0.09	&	1.97	$\pm$	0.21	\\
9	&	--	&	--	&	--	&	19	$\pm$	0.3	&	3.95	$\pm$	0.02	&	3.03	$\pm$	0.05	&	8.9	$\pm$	0.3	&	3.82	$\pm$	0.04	&	2.5	$\pm$	0.1	\\
10	&	--	&	--	&	--	&	5.9	$\pm$	0.2	&	-8.59	$\pm$	0.09	&	5.56	$\pm$	0.22	&	3.6	$\pm$	0.2	&	-8.46	$\pm$	0.16	&	5.52	$\pm$	0.37	\\
\hline\noalign{\smallskip}
\end{tabular}
\end{center}
\end{table*}

\begin{table*}
\begin{center}
\caption[]{CH Column Densities during Pulsar-OFF and CH Column Density Upper Limits within the Pulsar Beam.}
\label{tab:ch_N_off}
\setlength{\tabcolsep}{1.9pt}
\begin{tabular}{c|c|c|c|c|c|c}
\hline\noalign{\smallskip}
&\multicolumn{3}{c|}{During Pulsar-OFF}&\multicolumn{3}{c}{Upper Limits within the Pulsar Beam}\\
\hline\noalign{\smallskip}
   & 3264 MHz &3335 MHz  &3349 MHz& 3264 MHz &3335 MHz  &3349 MHz\\
\hline\noalign{\smallskip}
 Features   &$N(\mathrm{CH})_{\rm OFF}$& $N(\mathrm{CH})_{\rm OFF}$ & $N(\mathrm{CH})_{\rm OFF}$ &$N(\rm CH)_{\rm rms}$& $N(\rm CH)_{\rm rms}$ & $N(\rm CH)_{\rm rms}$ \\
 (No.)    & ($10^{13}$cm$^{\rm -2}$)& ($10^{13}$cm$^{\rm -2}$) & ($10^{13}$cm$^{\rm -2}$) & ($10^{13}$cm$^{\rm -2}$)& ($10^{13}$cm$^{\rm -2}$) & ($10^{13}$cm$^{\rm -2}$)   \\
\hline\noalign{\smallskip}
1	&	[0.2, 0.3]	&	[0.3, 0.4]	&	[0.4, 0.5]& -- & -- & --	\\
2	&	[4.4, 7.2]	&	[3.4, 5.5]	&	[6.6, 8.7]& -- & -- & --		\\
3	&	[0.7, 1.1]	&	[1.0, 1.7]	&	[1.8, 2.4]& -- & -- & --		\\
4	&	[2.3, 3.7]	&	[1.7, 2.7]	&	[2.9, 3.8]& -- & -- & --		\\
5	&	[4.8, 7.9]	&	[3.9, 6.4]	&	[7.4, 9.8]& -- & -- & --		\\
6	&	[2.1, 3.5]	&	[1.6, 2.7]	&	[3.2, 4.2]&	3.7	&	1.4	&	9.8	\\
7	&	--	&	[1.5, 2.5]	&	[2.1, 2.8]&	-- &	0.6	&	0.3	\\
8	&	[0.05, 0.1]	&	[0.3, 0.4]	&	[0.3, 0.5]&	1.8	&	0.7	&	0.8		\\
9	&	       --	&	[0.5, 0.8]	&	[0.7, 0.9]&	--	&	0.3	&	4.9	\\
10	&       	--	&	[0.3, 0.4]	&	[0.6, 0.8]&	--	&	0.9	&	3.0		\\
\hline\noalign{\smallskip}
\end{tabular}
\end{center}
\end{table*}

\section{Conclusions}
\label{sect:concl}

Using the UWL on Murriyang, CSIRO's Parkes Radio Telescope, we conducted the first search for potential pulsed CH maser emission stimulated by PSR~J1644$-$4559, with a variability timescale of 78 ms, at 3264, 3335, and 3349 MHz. 
We detected ten CH emission features at 3335 and 3349 MHz, and seven features at 3264 MHz (limited by the frequency cutoff at the edge of subband 20), during both pulsar-ON and pulsar-OFF phases. 
The observed velocities of these features are consistent with the OH emission and absorption reported by \citet{Weisberg_etal_2005}. 
We estimated the CH column density for clouds within the Parkes beam to range from 0.05 to 9.8$\times 10^{13}$ cm$^{-2}$. 
Adopting a typical CH abundance relative to $\mathrm{H}_2$, we infer that these clouds are likely in diffuse and translucent states. 
We derived the upper limits for the CH column density within the pulsar beam, ranging from 0.3 to 9.8$\times 10^{13}$ cm$^{-2}$. 
A comparison of the column densities within the pulsar beam and the Parkes beam suggests that CH clouds may exhibit clumpiness and substructure in certain regions. 
No significant maser emission was detected in the optical depth spectra. 
As a natural byproduct of our search for pulsed stimulated emission, we also searched for absorption of the pulsar signal by CH, and found none, as expected from astrophysical considerations. 
The upper limits for the maser amplification factors of the three 3.3 GHz CH lines toward PSR~J1644$-$4559 at 3264, 3335, and 3349 MHz are 1.014, 1.009, and 1.009, respectively.

\normalem
\begin{acknowledgements}
This work is supported by the National Natural Science Foundation of China (NSFC) program No.\ 11988101,  12203044, 12473023, 
by the Leading Innovation and Entrepreneurship Team of Zhejiang Province of China grant No.\ 2023R01008, by Key R\&D Program of Zhejiang grant No.\ 2024SSYS0012, 
and by the University Annual Scientific Research Plan of Anhui Province (No. 2023AH030052, No. 2022AH010013),  Cultivation Project for FAST Scientific Payoff and Research Achievement of CAMS-CAS. 
D.X. acknowledges the support of the Natural Sciences and Engineering Research Council of Canada (NSERC), [funding reference number 568580]. 
D.X. also acknowledges support from the Eric and Wendy Schmidt AI in Science Postdoctoral Fellowship Program, a program of Schmidt Sciences.
Murriyang, CSIRO's Parkes radio telescope, is part of the Australia Telescope National Facility (https://ror.org/05qajvd42) which is funded by the Australian Government for operation as a National Facility managed by CSIRO. We acknowledge the Wiradjuri people as the Traditional Owners of the Observatory site. 
We express our thanks to Lawrence Toomey, Andrew Jameson, Willem van Straten, James A. Green, and Stacy Mader for help with observations and data processing.

\end{acknowledgements}
  
\bibliographystyle{raa}
\bibliography{bibtex}

\begin{thebibliography}{34}
\providecommand\natexlab[1]{#1}
\providecommand\JournalTitle[1]{#1}

\bibitem[{Black} \& {Dalgarno}(1973)]{1973ApL....15...79B}
{Black}, J.~H., \& {Dalgarno}, A. 1973, \aplett, 15, 79

\bibitem[{Black} {et~al.}(1975)]{1975ApJ...199..633B}
{Black}, J.~H., {Dalgarno}, A., \& {Oppenheimer}, M. 1975, \apj, 199, 633

\bibitem[{Calabretta} {et~al.}(2014)]{CHIPASS_2014}
{Calabretta}, M.~R., {Staveley-Smith}, L., \& {Barnes}, D.~G. 2014, \pasa, 31,
  e007

\bibitem[{Carretti} {et~al.}(2019)]{SPASS_2019}
{Carretti}, E., {Haverkorn}, M., {Staveley-Smith}, L., {et~al.} 2019, \mnras,
  489, 2330

\bibitem[{Clifton} {et~al.}(1988)]{Clifton_1988_HI_PSR}
{Clifton}, T.~R., {Frail}, D.~A., {Kulkarni}, S.~R., \& {Weisberg}, J.~M. 1988,
  \apj, 333, 332

\bibitem[{Dawson} {et~al.}(2022)]{SPLASH_2022}
{Dawson}, J.~R., {Jones}, P.~A., {Purcell}, C., {et~al.} 2022, \mnras, 512,
  3345

\bibitem[{Draine}(2011)]{2011piim.book.....D}
{Draine}, B.~T. 2011, {Physics of the Interstellar and Intergalactic Medium}

\bibitem[{Elitzur}(1992)]{1992ASSL..170.....E}
{Elitzur}, M. 1992, {Astronomical masers}, Vol. 170

\bibitem[{Frail} {et~al.}(1994)]{Frail_1994_PSR_HI}
{Frail}, D.~A., {Weisberg}, J.~M., {Cordes}, J.~M., \& {Mathers}, C. 1994,
  \apj, 436, 144

\bibitem[{Genzel} {et~al.}(1979)]{1979A&A....73..253G}
{Genzel}, R., {Downes}, D., {Pauls}, T., {Wilson}, T.~L., \& {Bieging}, J.
  1979, \aap, 73, 253

\bibitem[{Hobbs} {et~al.}(2020)]{Hobbs_UWL_2020}
{Hobbs}, G., {Manchester}, R.~N., {Dunning}, A., {et~al.} 2020, \pasa, 37, e012

\bibitem[{Jacob}(2023)]{Jacob_CH_2023}
{Jacob}, A.~M. 2023, \apss, 368, 76

\bibitem[{Jacob} {et~al.}(2021)]{Jacob_2021_CH}
{Jacob}, A.~M., {Menten}, K.~M., {Wiesemeyer}, H., \& {Ortiz-Le{\'o}n}, G.~N.
  2021, \aap, 650, A133

\bibitem[{Jacob} {et~al.}(2022)]{2022ApJ...930..141J}
{Jacob}, A.~M., {Neufeld}, D.~A., {Schilke}, P., {et~al.} 2022, \apj, 930, 141

\bibitem[{Johnston} {et~al.}(2003)]{Simon_2003_HI}
{Johnston}, S., {Koribalski}, B., {Wilson}, W., \& {Walker}, M. 2003, \mnras,
  341, 941

\bibitem[{Lewandowski} {et~al.}(2013)]{2013MNRAS.434...69L}
{Lewandowski}, W., {Dembska}, M., {Kijak}, J., \& {Kowali{\'n}ska}, M. 2013,
  \mnras, 434, 69

\bibitem[{Li} {et~al.}(2018)]{2018ApJS..235....1L}
{Li}, D., {Tang}, N., {Nguyen}, H., {et~al.} 2018, \apjs, 235, 1

\bibitem[{Liszt} \& {Lucas}(2002)]{2002A&A...391..693L}
{Liszt}, H., \& {Lucas}, R. 2002, \aap, 391, 693

\bibitem[{Liu} {et~al.}(2021)]{Liu_HI_2021}
{Liu}, M., {Kr{\v{c}}o}, M., {Li}, D., {et~al.} 2021, \apjl, 911, L13

\bibitem[{Minter}(2008)]{Minter_2008_OH_PSR}
{Minter}, A.~H. 2008, \apj, 677, 373

\bibitem[{Rydbeck} {et~al.}(1976)]{1976ApJS...31..333R}
{Rydbeck}, O.~E.~H., {Kollberg}, E., {Hjalmarson}, A., {et~al.} 1976, \apjs,
  31, 333

\bibitem[{Sheffer} {et~al.}(2008)]{2008ApJ...687.1075S}
{Sheffer}, Y., {Rogers}, M., {Federman}, S.~R., {et~al.} 2008, \apj, 687, 1075

\bibitem[{Stanimirovi{\'c}} {et~al.}(2003)]{Snez_OH_2003}
{Stanimirovi{\'c}}, S., {Weisberg}, J.~M., {Dickey}, J.~M., {et~al.} 2003,
  \apj, 592, 953

\bibitem[{Stanimirovi{\'c}} {et~al.}(2010)]{Snez_HI_2010}
{Stanimirovi{\'c}}, S., {Weisberg}, J.~M., {Pei}, Z., {Tuttle}, K., \& {Green},
  J.~T. 2010, \apj, 720, 415

\bibitem[{Suutarinen} {et~al.}(2011)]{Suutarinen_TMC1_CH_2011}
{Suutarinen}, A., {Geppert}, W.~D., {Harju}, J., {et~al.} 2011, \aap, 531, A121

\bibitem[{Tang} {et~al.}(2021)]{Ningyu_2021_CH}
{Tang}, N., {Li}, D., {Luo}, G., {et~al.} 2021, \apjs, 257, 47

\bibitem[{Turner}(1988)]{Turner_1988_CH}
{Turner}, B.~E. 1988, \apj, 329, 425

\bibitem[{van Straten} \& {Bailes}(2011)]{Willem_dspsr_2011}
{van Straten}, W., \& {Bailes}, M. 2011, \pasa, 28, 1

\bibitem[{van Straten} {et~al.}(2012)]{Willem_psrchive_2012}
{van Straten}, W., {Demorest}, P., \& {Oslowski}, S. 2012, Astronomical
  Research and Technology, 9, 237

\bibitem[{Verbiest} {et~al.}(2012)]{Verbiest_PSR_dist_2012}
{Verbiest}, J.~P.~W., {Weisberg}, J.~M., {Chael}, A.~A., {Lee}, K.~J., \&
  {Lorimer}, D.~R. 2012, \apj, 755, 39

\bibitem[{Weisberg} {et~al.}(1979)]{Weisberg_etal_1979}
{Weisberg}, J.~M., {Boriakoff}, V., \& {Rankin}, J. 1979, \aap, 77, 204

\bibitem[{Weisberg} {et~al.}(2005)]{Weisberg_etal_2005}
{Weisberg}, J.~M., {Johnston}, S., {Koribalski}, B., \& {Stanimirovi{\'c}}, S.
  2005, Science, 309, 106

\bibitem[{Xu} \& {Li}(2016)]{Xu_2016_CH}
{Xu}, D., \& {Li}, D. 2016, \apj, 833, 90

\bibitem[{Xu} {et~al.}(2016)]{Xu_2016_OH}
{Xu}, D., {Li}, D., {Yue}, N., \& {Goldsmith}, P.~F. 2016, \apj, 819, 22

\end{thebibliography}

\end{document}